\documentclass[
  shortnames]{jss}

\usepackage{orcidlink,thumbpdf,lmodern}

\usepackage[utf8]{inputenc}

\author{
H. Sherry Zhang\\Monash University \And Dianne Cook\\Monash University \And Ursula Laa\\University of Natural \\ Resources and Life Sciences \AND Nicolas Langrené\\BNU-HKBU \\ United International College \And Patricia Menéndez\\University of Melbourne
}
\title{\pkg{cubble}: An \proglang{R} Package for Organizing and Wrangling Multivariate Spatio-temporal Data}

\Plainauthor{H. Sherry Zhang, Dianne Cook, Ursula Laa, Nicolas Langrené, Patricia Menéndez}
\Plaintitle{cubble: An R Package for Organizing and Wrangling Multivariate Spatio-temporal Data}

\Abstract{
Multivariate spatio-temporal data refers to multiple measurements taken across space and time. For many analyses, spatial and time components can be separately studied: for example, to explore the temporal trend of one variable for a single spatial location, or to model the spatial distribution of one variable at a given time. However for some studies, it is important to analyze different aspects of the spatio-temporal data simultaneously, for instance, temporal trends of multiple variables across locations. In order to facilitate the study of different portions or combinations of spatio-temporal data, we introduce a new class, \code{cubble}, with a suite of functions enabling easy slicing and dicing on different spatio-temporal components. The proposed \code{cubble} class ensures that all the components of the data are easy to access and manipulate while providing flexibility for data analysis. In addition, the \pkg{cubble} package facilitates visual and numerical explorations of the data while easing data wrangling and modelling. The \code{cubble} class and the tools implemented in the package are illustrated with examples from climate data analysis.
}

\Plainkeywords{spatial, temporal, spatio-temporal, R, environmental data, exploratory data analysis}

\Address{
    H. Sherry Zhang\\
    Monash University\\
    21 Chancellors Walk, Clayton VIC 3800 Australia\\
  E-mail: \email{huize.zhang@monash.edu}\\
  
      Dianne Cook\\
    Monash University\\
    21 Chancellors Walk, Clayton VIC 3800 Australia\\
  E-mail: \href{mailto:dicook@monash.edu}{\nolinkurl{dicook@monash.edu}}\\
  
      Ursula Laa\\
    University of Natural \\ Resources and Life Sciences\\
    Gregor-Mendel-Straße 33, 1180 Wien, Austria\\
  E-mail: \href{mailto:ursula.laa@boku.ac.at}{\nolinkurl{ursula.laa@boku.ac.at}}\\
  
      Nicolas Langrené\\
    BNU-HKBU \\ United International College\\
    2000 Jintong Road, Tangjiawan, Zhuhai, Guangdong Province, China\\
  E-mail: \href{mailto:nicolaslangrene@uic.edu.cn}{\nolinkurl{nicolaslangrene@uic.edu.cn}}\\
  
      Patricia Menéndez\\
    Monash University\\
    813 Swanston Street, Parkville VIC 3052, Australia\\
  E-mail: \href{mailto:Patricia.menendez@unimelb.edu.au}{\nolinkurl{Patricia.menendez@unimelb.edu.au}}\\
  
  }

\providecommand{\tightlist}{%
  \setlength{\itemsep}{0pt}\setlength{\parskip}{0pt}}

\usepackage{longtable,booktabs,array}
\usepackage{calc} 
\usepackage{etoolbox}
\makeatletter
\patchcmd\longtable{\par}{\if@noskipsec\mbox{}\fi\par}{}{}
\makeatother
\IfFileExists{footnotehyper.sty}{\usepackage{footnotehyper}}{\usepackage{footnote}}
\makesavenoteenv{longtable}

\usepackage{amsmath} \usepackage{array}

\begin{document}

\hypertarget{introduction}{%
\section{Introduction}\label{introduction}}

Spatio-temporal data \citep{bivand2008applied, lovelace2019geocomputation, pebesma2019spatial} has a spatial component referring to the location of each observation and a temporal component that is recorded at regular or irregular time intervals. It may also include multiple variables measured at each spatial and temporal values. With spatio-temporal data, one can fix the time to explore the spatial features of the data, fix the spatial location/s to explore temporal aspects, or dynamically explore the space and time simultaneously. In order to computationally explore the spatial, temporal and spatio-temporal and multiple variable aspects of such data, it needs to be stored in a data object that allows the user to query, group and dissect all the different data faces.

The Comprehensive \proglang{R} Archive Network (CRAN) task view SpatioTemporal \citep{ctvspatiotemporal} gathers information about \proglang{R} packages designed for spatio-temporal data and it has a section on \emph{Representing data} that lists existing spatio-temporal data representations used in \proglang{R}. Among them, the \pkg{spacetime} package \citep{spacetime} implements four S4 classes to handle spatio-temporal data with different spatio-temporal layouts (full grid, sparse grid, irregular, and trajectory). The \pkg{stars} package \citep{stars} implements an S3 class built from dense arrays.

However, the data representation implemented in those packages might present certain challenges when applying the principles of tidy data \citep{tidydata} for data analysis. The concept of tidy data is based on three principles regarding how data should be organized in tables to facilitate easier analysis: 1) one observation a row, 2) one variable a column, and 3) one type of observation a table. The third principle of tidy data is particularly relevant for spatio-temporal data since these data are naturally observed at different units: the spatial locations and the temporal units. While the tidyverse suite of R packages implements data wrangling and visualization tools primarily focused on working with single tables, there are not many tools available for handling relational data specifically for spatio-temporal data. This motivates a new design to organise spatio-temporal data in a way that would make data wrangling, visualizing and analyzing easier.

This paper presents the \proglang{R} package, \pkg{cubble}, which implements a new cubble class to organize spatial and temporal variables as two forms of a single data object so that they can be wrangled separately or combined, while being kept synchronized. Among the four spacetime layouts in \citet{spacetime}, the \code{cubble} class can handle the full grid layout and the sparse grid layout. The software is available from the Comprehensive \proglang{R} Archive Network (CRAN) at \url{https://CRAN.R-project.org/package=cubble}.

The rest of the paper is organized as follows: Section \ref{cubble} presents the main design and functionality of the \pkg{cubble} package. Section \ref{others} explains how the \pkg{cubble} package deals with more advanced considerations, including data matching and how the package fits with existing static and interactive visualization tools. Moreover we also illustrate how the \pkg{cubble} package deals with spatio-temporal data transformations. Section \ref{examples} uses primarily Australian weather station data as examples to demonstrate the use of the package. An example of how the \pkg{cubble} package handles Network Common Data Form (NetCDF) data is also provided. Section \ref{conclude} discusses the paper contributions and future directions.

\hypertarget{cubble}{%
\section{The cubble package}\label{cubble}}

The cubble class includes two subclasses: the spatial cubble and the temporal cubble, which can be pivoted back and forth to focus on the two aspects of the spatio-temporal data, as illustrated in Figure \ref{fig:face}. This section provides an overview of the \pkg{cubble} package, including the cubble class and its attributes, class creation and coercion, a summary of implemented functionality, the compatibility with other spatial and temporal packages (\pkg{sf} and \pkg{tsibble}), and a comparison with other spatio-temporal packages (\pkg{stars} and \pkg{sftime}).

\hypertarget{sec-cb-object}{%
\subsection{The cubble class}\label{sec-cb-object}}

The cubble class is an S3 class built on tibble that allows the spatio-temporal data to be wrangled in two forms (subclasses):

\begin{itemize}
\tightlist
\item
  a spatial cubble with class \code{c("spatial_cubble_df", "cubble_df")}
\item
  a temporal cubble with class \code{c("temporal_cubble_df", "cubble_df")}
\end{itemize}

In a spatial cubble object, spatial variables are organised as columns and temporal variables are nested within a specialised \code{ts} column. For example, the spatial cubble object, \code{cb_spatial} printed below, contains weather records of three airport stations from the Global Historical Climatology Network Daily (GHCND) database \citep{menne2012overview}. In this case, the spatial cubble is convenient for wrangling the spatial variables:

\begin{CodeChunk}
\begin{CodeInput}
R> cb_spatial
\end{CodeInput}
\begin{CodeOutput}
# cubble:   key: id [3], index: date, nested form
# spatial:  [144.8321, -37.98, 145.0964, -37.6655], Missing CRS!
# temporal: date [date], prcp [dbl], tmax [dbl], tmin [dbl]
  id           long   lat  elev name              wmo_id ts               
  <chr>       <dbl> <dbl> <dbl> <chr>              <dbl> <list>           
1 ASN00086038  145. -37.7  78.4 essendon airport   95866 <tibble [10 x 4]>
2 ASN00086077  145. -38.0  12.1 moorabbin airport  94870 <tibble [10 x 4]>
3 ASN00086282  145. -37.7 113.  melbourne airport  94866 <tibble [10 x 4]>
\end{CodeOutput}
\end{CodeChunk}

In a temporal cubble, temporal variables are expanded in the long form and spatial variables are stored as a data attribute. The temporal cubble object, \code{cb_temporal}, contains the same spatio-temporal data as the spatial cubble object, \code{cb_spatial}, but in a structure that is easier for temporal analysis:

\begin{CodeChunk}
\begin{CodeInput}
R> cb_temporal
\end{CodeInput}
\begin{CodeOutput}
# cubble:   key: id [3], index: date, long form
# temporal: 2020-01-01 -- 2020-01-10 [1D], no gaps
# spatial:  long [dbl], lat [dbl], elev [dbl], name [chr], wmo_id [dbl]
  id          date        prcp  tmax  tmin
  <chr>       <date>     <dbl> <dbl> <dbl>
1 ASN00086038 2020-01-01     0  26.8  11  
2 ASN00086038 2020-01-02     0  26.3  12.2
3 ASN00086038 2020-01-03     0  34.5  12.7
4 ASN00086038 2020-01-04     0  29.3  18.8
5 ASN00086038 2020-01-05    18  16.1  12.5
# i 25 more rows
\end{CodeOutput}
\end{CodeChunk}

\hypertarget{the-cubble-attributes}{%
\subsubsection{The cubble attributes}\label{the-cubble-attributes}}

Both cubble objects inherit tibble's attributes (which originates from data frames): \code{class}, \code{row.names}, and \code{names}. Additionally, both have three specialised attributes: \code{key}, \code{index}, and \code{coords}, where \code{key} and \code{index} are used as introduced in the \pkg{tsibble} package \citep{tsibble}. In cubble, the \code{key} attribute identifies the row in the spatial cubble (given the internal use of \code{tidyr::nest()} for nesting), and when combined with the \code{index} argument, it identifies the row in the temporal cubble. Currently, cubble only supports one variable as the key. The accepted temporal classes for \code{index} includes the base R classes \code{Date}, \code{POSIXlt}, \code{POSIXct}, as well as tsibble's \code{yearmonth}, \code{yearweek}, and \code{yearquarter} classes. The \code{coords} attribute represents an ordered pair of coordinates that can be either an unprojected pair of longitude and latitude, or a projected easting and northing value. Moreover, temporal cubbles have a special attribute called \code{spatial} to store the spatial variables. Shortcut functions are available to extract attributes from the temporal cubble object, for example, \code{spatial()} for extracting spatial variables:

\begin{CodeChunk}
\begin{CodeInput}
R> spatial(cb_temporal)
\end{CodeInput}
\begin{CodeOutput}
# A tibble: 3 x 6
  id           long   lat  elev name              wmo_id
  <chr>       <dbl> <dbl> <dbl> <chr>              <dbl>
1 ASN00086038  145. -37.7  78.4 essendon airport   95866
2 ASN00086077  145. -38.0  12.1 moorabbin airport  94870
3 ASN00086282  145. -37.7 113.  melbourne airport  94866
\end{CodeOutput}
\end{CodeChunk}

\hypertarget{create}{%
\subsection{Creation and coercion}\label{create}}

The spatial and temporal aspect of spatio-temporal data are often stored separately in the database. For climate data, analysts may initially receive station metadata and then query the time series based on the metadata. A (spatial) cubble object can be constructed from separate spatial and temporal tables using the function \code{make_cubble()}. The three attributes \code{key}, \code{index}, and \code{coords} need to be specified. The following code creates a spatial cubble from its spatial component, \code{stations} and temporal component \code{meteo}:

\begin{CodeChunk}
\begin{CodeInput}
R> stations
\end{CodeInput}
\begin{CodeOutput}
# A tibble: 3 x 6
  id           long   lat  elev name              wmo_id
  <chr>       <dbl> <dbl> <dbl> <chr>              <dbl>
1 ASN00086038  145. -37.7  78.4 essendon airport   95866
2 ASN00086077  145. -38.0  12.1 moorabbin airport  94870
3 ASN00086282  145. -37.7 113.  melbourne airport  94866
\end{CodeOutput}
\begin{CodeInput}
R> meteo
\end{CodeInput}
\begin{CodeOutput}
# A tibble: 30 x 5
  id          date        prcp  tmax  tmin
  <chr>       <date>     <dbl> <dbl> <dbl>
1 ASN00086038 2020-01-01     0  26.8  11  
2 ASN00086038 2020-01-02     0  26.3  12.2
3 ASN00086038 2020-01-03     0  34.5  12.7
4 ASN00086038 2020-01-04     0  29.3  18.8
5 ASN00086038 2020-01-05    18  16.1  12.5
# i 25 more rows
\end{CodeOutput}
\begin{CodeInput}
R> make_cubble(spatial = stations, temporal = meteo,
+             key = id, index = date, coords = c(long, lat))
\end{CodeInput}
\begin{CodeOutput}
# cubble:   key: id [3], index: date, nested form
# spatial:  [144.8321, -37.98, 145.0964, -37.6655], Missing CRS!
# temporal: date [date], prcp [dbl], tmax [dbl], tmin [dbl]
  id           long   lat  elev name              wmo_id ts               
  <chr>       <dbl> <dbl> <dbl> <chr>              <dbl> <list>           
1 ASN00086038  145. -37.7  78.4 essendon airport   95866 <tibble [10 x 4]>
2 ASN00086077  145. -38.0  12.1 moorabbin airport  94870 <tibble [10 x 4]>
3 ASN00086282  145. -37.7 113.  melbourne airport  94866 <tibble [10 x 4]>
\end{CodeOutput}
\end{CodeChunk}

Other \proglang{R} spatio-temporal objects can be coerced into a \code{cubble} object with the function \code{as_cubble()}. This includes a joined \code{tibble} or \code{data.frame} object, a NetCDF object, a \code{stars} object \citep{stars}, and a \code{sftime} object \citep{sftime}. In the example below, the spatial cubble object is created from \code{climate_flat}, which combines the previous \code{stations} and \code{meteo} into a single tibble object:

\begin{CodeChunk}
\begin{CodeInput}
R> climate_flat |> as_cubble(key = id, index = date, coords = c(long, lat))
\end{CodeInput}
\begin{CodeOutput}
# cubble:   key: id [3], index: date, nested form
# spatial:  [144.8321, -37.98, 145.0964, -37.6655], Missing CRS!
# temporal: date [date], prcp [dbl], tmax [dbl], tmin [dbl]
  id           long   lat  elev name              wmo_id ts               
  <chr>       <dbl> <dbl> <dbl> <chr>              <dbl> <list>           
1 ASN00086038  145. -37.7  78.4 essendon airport   95866 <tibble [10 x 4]>
2 ASN00086077  145. -38.0  12.1 moorabbin airport  94870 <tibble [10 x 4]>
3 ASN00086282  145. -37.7 113.  melbourne airport  94866 <tibble [10 x 4]>
\end{CodeOutput}
\end{CodeChunk}

\hypertarget{functions-and-methods}{%
\subsection{Functions and methods}\label{functions-and-methods}}

The \pkg{cubble} package has several functions implemented for data wrangling and to facilitate data analysis as summarized in Table \ref{tab:funs}. In addition, for each of the three cubble classes there are a number of methods implemented that facilitates the handling of the data as shown in Table \ref{tab:methods}. In particular, the \code{cubble_df} class handles methods that behave consistently in both spatial and temporal cubble. When the method works differently internally on the spatial and temporal cubble, it is implemented separately in \code{spatial_cubble_df} and \code{temporal_cubble_df}.

\begin{table}
\begin{tabular}{p{0.1\textwidth}p{0.8\textwidth}}
\hline
Category & Functions \\
\hline
base R & \texttt{{[}}, \texttt{{[}{[}\textless{}-}, \texttt{names\textless{}-} \\
tidyverse & \texttt{dplyr\_row\_slice}, \texttt{dplyr\_col\_modify}, \texttt{dplyr\_reconstruct}, \texttt{select}, \texttt{mutate}, \texttt{arrange}, \texttt{filter}, \texttt{group\_by}, \texttt{ungroup}, \texttt{summarise}, \texttt{select}, \texttt{slice}, \texttt{rowwise}, \texttt{rename}, \texttt{bind\_rows}, \texttt{bind\_cols}, \texttt{relocate}, \texttt{type\_sum}, the slice family (\texttt{slice\_head}, \texttt{slice\_tail}, \texttt{slice\_max}, \texttt{slice\_min}, \texttt{slice\_sample}) and the join family (\texttt{left\_join}, \texttt{right\_join}, \texttt{inner\_join}, \texttt{full\_join}, \texttt{anti\_join}, \texttt{semi\_join}) \\
cubble & \texttt{as\_cubble}, \texttt{cubble}, \texttt{make\_cubble}, \texttt{check\_key}, \texttt{face\_temporal}, \texttt{face\_spatial}, \texttt{unfold}, \texttt{key}, \texttt{key\_vars}, \texttt{key\_data}, \texttt{index}, \texttt{index\_var}, \texttt{coords}, \texttt{spatial}, \texttt{match\_sites}, \texttt{match\_spatial}, \texttt{match\_temporal}, \texttt{geom\_glyph}, \texttt{geom\_glyph\_box}, \texttt{geom\_glyph\_line}, \texttt{make\_spatial\_sf}, \texttt{make\_temporal\_tsibble}, \texttt{fill\_gaps}, and \texttt{scan\_gaps} \\
\hline
\end{tabular}
\caption{An overview of functions implemented in the cubble package, categorised into base R, tidyverse, and cubble functions.}
\label{tab:funs}
\end{table}

\begin{table}
\begin{tabular}{p{0.25\textwidth}p{0.65\textwidth}}
\hline
Class & Method \\
\hline
\texttt{cubble\_df} & \texttt{{[}{[}\textless{}-,\ dplyr\_col\_modify,\ key\_data,\ key\_vars,\ key,\ print} \\
\texttt{spatial\_cubble\_df} & \texttt{{[},\ names\textless{}-,\ \ tbl\_sum,\ dplyr\_reconstruct,\ dplyr\_row\_slice,\ face\_spatial,\ face\_temporal,\ unfold,\ arrange,\ rename,\ rowwise,\ group\_by,\ ungroup,\ select,\ spatial,\ summarise,\ unfold,\ update\_cubble} \\
\texttt{temporal\_cubble\_df} & \texttt{{[},\ names\textless{}-,\ tbl\_sum,\ arrange,\ dplyr\_reconstruct,\ dplyr\_row\_slice,\ face\_spatial,\ face\_temporal,\ unfold,\ fill\_gaps,\ group\_by,\ ungroup,\ \ rename,\ rowwise,\ scan\_gaps,\ select,\ spatial,\ summarise,\ tbl\_sum,\ bind\_rows,\ bind\_cols,\ update\_cubble} \\
\hline
\end{tabular}
\caption{An overview of the methods implemented in the three \code{cubble} classes. Methods are implemented in the \code{cubble\_df} class when they behave consistently across the spatial and temporal cubble; otherwise, they are implemented separately.}
\label{tab:methods} 
\end{table}

The pair of cubble verbs, \code{face_temporal()} and \code{face_spatial()}, pivots the cubble object between its two forms or faces, as illustrated in Figure \ref{fig:face}. The code applies \code{face_temporal()} on the spatial cubble, \code{cb_spatial}, introduced in Section \ref{cb-object} to get a temporal cubble:

\begin{CodeChunk}
\begin{CodeInput}
R> face_temporal(cb_spatial)
\end{CodeInput}
\begin{CodeOutput}
# cubble:   key: id [3], index: date, long form
# temporal: 2020-01-01 -- 2020-01-10 [1D], no gaps
# spatial:  long [dbl], lat [dbl], elev [dbl], name [chr], wmo_id [dbl]
  id          date        prcp  tmax  tmin
  <chr>       <date>     <dbl> <dbl> <dbl>
1 ASN00086038 2020-01-01     0  26.8  11  
2 ASN00086038 2020-01-02     0  26.3  12.2
3 ASN00086038 2020-01-03     0  34.5  12.7
4 ASN00086038 2020-01-04     0  29.3  18.8
5 ASN00086038 2020-01-05    18  16.1  12.5
# i 25 more rows
\end{CodeOutput}
\end{CodeChunk}

Both verbs are the exact inverse of each other and apply both functions on a cubble object will result in the object itself:

\begin{CodeChunk}
\begin{CodeInput}
R> face_spatial(face_temporal(cb_spatial))
\end{CodeInput}
\begin{CodeOutput}
# cubble:   key: id [3], index: date, nested form
# spatial:  [144.8321, -37.98, 145.0964, -37.6655], Missing CRS!
# temporal: date [date], prcp [dbl], tmax [dbl], tmin [dbl]
  id           long   lat  elev name              wmo_id ts               
  <chr>       <dbl> <dbl> <dbl> <chr>              <dbl> <list>           
1 ASN00086038  145. -37.7  78.4 essendon airport   95866 <tibble [10 x 4]>
2 ASN00086077  145. -38.0  12.1 moorabbin airport  94870 <tibble [10 x 4]>
3 ASN00086282  145. -37.7 113.  melbourne airport  94866 <tibble [10 x 4]>
\end{CodeOutput}
\end{CodeChunk}

\begin{CodeChunk}
\begin{figure}

{\centering \includegraphics[width=1\linewidth]{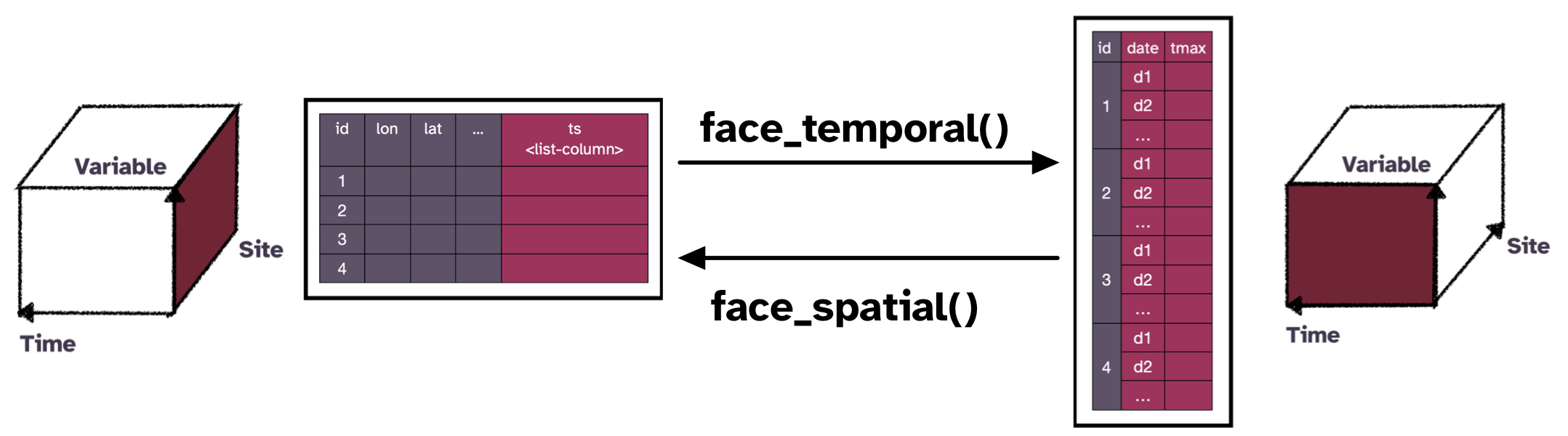} 

}

\caption{Illustration of main functions. To focus on the temporal variables \code{face\_temporal()} converts a spatial cubble into a temporal cubble. To focus on the spatial variables \code{face\_spatial()} transforms a temporal cubble into a spatial cubble. This pivoting makes it easy to separately do spatial or temporal analysis.}\label{fig:face}
\end{figure}
\end{CodeChunk}

To enable operations involve both spatial and temporal variables, the function \code{unfold} incorporates spatial variables into the temporal cubble. Below is an example to include the coordinate columns (\code{long} and \code{lat}) into \code{cb_temporal} to prepare the data for a glyph map transformation, which will be discussed in Section \ref{st_transformation}.

\begin{CodeChunk}
\begin{CodeInput}
R> cb_temporal |> unfold(long, lat)
\end{CodeInput}
\begin{CodeOutput}
# cubble:   key: id [3], index: date, long form
# temporal: 2020-01-01 -- 2020-01-10 [1D], no gaps
# spatial:  long [dbl], lat [dbl], elev [dbl], name [chr], wmo_id [dbl]
  id          date        prcp  tmax  tmin  long   lat
  <chr>       <date>     <dbl> <dbl> <dbl> <dbl> <dbl>
1 ASN00086038 2020-01-01     0  26.8  11    145. -37.7
2 ASN00086038 2020-01-02     0  26.3  12.2  145. -37.7
3 ASN00086038 2020-01-03     0  34.5  12.7  145. -37.7
4 ASN00086038 2020-01-04     0  29.3  18.8  145. -37.7
5 ASN00086038 2020-01-05    18  16.1  12.5  145. -37.7
# i 25 more rows
\end{CodeOutput}
\end{CodeChunk}

\hypertarget{compatibility-with-tsibble-and-sf}{%
\subsection{Compatibility with tsibble and sf}\label{compatibility-with-tsibble-and-sf}}

Analysts often have their preferred spatial or temporal data structure for spatial or temporal analysis, which they may wish to continue using for spatio-temporal analysis. With \code{cubble}, analysts can incorporate the \code{tsibble} class in a temporal cubble and the \code{sf} class in a spatial cubble.

\hypertarget{using-a-tsibble-object-as-the-temporal-component}{%
\subsubsection{Using a tsibble object as the temporal component}\label{using-a-tsibble-object-as-the-temporal-component}}

The \code{key} and \code{index} arguments in a \code{cubble} object corresponds to the tsibble counterparts and they can be safely omitted, if the temporal component is a tsibble object (\code{tbl_ts}). The \code{tsibble} class (\code{tbl_ts}) from the input will be carried over to the temporal cubble, indicated by the \code{[tsibble]} in the header and in the object class:

\begin{CodeChunk}
\begin{CodeInput}
R> class(meteo_ts)
\end{CodeInput}
\begin{CodeOutput}
[1] "tbl_ts"     "tbl_df"     "tbl"        "data.frame"
\end{CodeOutput}
\begin{CodeInput}
R> ts_spatial <- make_cubble(
+   spatial = stations, temporal = meteo_ts, coords = c(long, lat))
R> (ts_temporal <- face_temporal(ts_spatial))
\end{CodeInput}
\begin{CodeOutput}
# cubble:   key: id [3], index: date, long form, [tsibble]
# temporal: 2020-01-01 -- 2020-01-10 [1D], no gaps
# spatial:  long [dbl], lat [dbl], elev [dbl], name [chr], wmo_id [dbl]
  id          date        prcp  tmax  tmin
  <chr>       <date>     <dbl> <dbl> <dbl>
1 ASN00086038 2020-01-01     0  26.8  11  
2 ASN00086038 2020-01-02     0  26.3  12.2
3 ASN00086038 2020-01-03     0  34.5  12.7
4 ASN00086038 2020-01-04     0  29.3  18.8
5 ASN00086038 2020-01-05    18  16.1  12.5
# i 25 more rows
\end{CodeOutput}
\begin{CodeInput}
R> class(ts_temporal)
\end{CodeInput}
\begin{CodeOutput}
[1] "temporal_cubble_df" "cubble_df"          "tbl_ts"            
[4] "tbl_df"             "tbl"                "data.frame"        
\end{CodeOutput}
\end{CodeChunk}

Methods applied to tsibble objects (\code{tbl_ts}) can also be applied to the temporal cubble objects, for example, checking whether the data contain temporal gaps:

\begin{CodeChunk}
\begin{CodeInput}
R> ts_temporal |> has_gaps()
\end{CodeInput}
\begin{CodeOutput}
# A tibble: 3 x 2
  id          .gaps
  <chr>       <lgl>
1 ASN00086038 FALSE
2 ASN00086077 FALSE
3 ASN00086282 FALSE
\end{CodeOutput}
\end{CodeChunk}

The temporal component of a created temporal cubble can include class \code{tbl_ts} to also be a tsibble object \code{tsibble} object using \code{make_temporal_tsibble()}. See the code example below using the \code{cb_temporal} object, created in Section \ref{create}:

\begin{CodeChunk}
\begin{CodeInput}
R> cb_temporal |> make_temporal_tsibble() 
\end{CodeInput}
\begin{CodeOutput}
# cubble:   key: id [3], index: date, long form, [tsibble]
# temporal: 2020-01-01 -- 2020-01-10 [1D], no gaps
# spatial:  long [dbl], lat [dbl], elev [dbl], name [chr], wmo_id [dbl]
  id          date        prcp  tmax  tmin
  <chr>       <date>     <dbl> <dbl> <dbl>
1 ASN00086038 2020-01-01     0  26.8  11  
2 ASN00086038 2020-01-02     0  26.3  12.2
3 ASN00086038 2020-01-03     0  34.5  12.7
4 ASN00086038 2020-01-04     0  29.3  18.8
5 ASN00086038 2020-01-05    18  16.1  12.5
# i 25 more rows
\end{CodeOutput}
\end{CodeChunk}

\hypertarget{using-an-sf-object-as-the-spatial-component}{%
\subsubsection{Using an sf object as the spatial component}\label{using-an-sf-object-as-the-spatial-component}}

Similarly, the spatial component of a cubble object can be an \code{sf} object and if the \code{coords} argument is omitted, it will be calculated from the \code{sf} geometry. The sf status is signalled by the \code{[sf]} label in the cubble header:

\begin{CodeChunk}
\begin{CodeInput}
R> (sf_spatial <- make_cubble(
+   spatial = stations_sf, temporal = meteo, 
+   key = id, index = date))
\end{CodeInput}
\begin{CodeOutput}
# cubble:   key: id [3], index: date, nested form, [sf]
# spatial:  [144.8321, -37.98, 145.0964, -37.6655], WGS 84
# temporal: date [date], prcp [dbl], tmax [dbl], tmin [dbl]
  id           elev name   wmo_id  long   lat            geometry ts      
  <chr>       <dbl> <chr>   <dbl> <dbl> <dbl>         <POINT [°]> <list>  
1 ASN00086038  78.4 essen~  95866  145. -37.7 (144.9066 -37.7276) <tibble>
2 ASN00086077  12.1 moora~  94870  145. -38.0   (145.0964 -37.98) <tibble>
3 ASN00086282 113.  melbo~  94866  145. -37.7 (144.8321 -37.6655) <tibble>
\end{CodeOutput}
\begin{CodeInput}
R> class(sf_spatial)
\end{CodeInput}
\begin{CodeOutput}
[1] "spatial_cubble_df" "cubble_df"         "sf"               
[4] "tbl_df"            "tbl"               "data.frame"       
\end{CodeOutput}
\end{CodeChunk}

This allows applying functions from the \code{sf} package to a cubble object, for example, to handle coordinate transformation with \code{st_transform()}:

\begin{CodeChunk}
\begin{CodeInput}
R> sf_spatial |> sf::st_transform(crs = "EPSG:3857")
\end{CodeInput}
\begin{CodeOutput}
# cubble:   key: id [3], index: date, nested form, [sf]
# spatial:  [16122635.6225205, -4576600.8687746, 16152057.3639371,
#   -4532279.35567565], WGS 84
# temporal: date [date], prcp [dbl], tmax [dbl], tmin [dbl]
  id           elev name   wmo_id  long   lat            geometry ts      
  <chr>       <dbl> <chr>   <dbl> <dbl> <dbl>         <POINT [°]> <list>  
1 ASN00086038  78.4 essen~  95866  145. -37.7 (16130929 -4541016) <tibble>
2 ASN00086077  12.1 moora~  94870  145. -38.0 (16152057 -4576601) <tibble>
3 ASN00086282 113.  melbo~  94866  145. -37.7 (16122636 -4532279) <tibble>
\end{CodeOutput}
\end{CodeChunk}

The spatial component of a created cubble can also be an \code{sf} object using \code{make_spatial_sf()}:

\begin{CodeChunk}
\begin{CodeInput}
R> cb_spatial |> make_spatial_sf() 
\end{CodeInput}
\begin{CodeOutput}
# cubble:   key: id [3], index: date, nested form, [sf]
# spatial:  [144.8321, -37.98, 145.0964, -37.6655], WGS 84
# temporal: date [date], prcp [dbl], tmax [dbl], tmin [dbl]
  id           long   lat  elev name   wmo_id ts                  geometry
  <chr>       <dbl> <dbl> <dbl> <chr>   <dbl> <list>           <POINT [°]>
1 ASN00086038  145. -37.7  78.4 essen~  95866 <tibble> (144.9066 -37.7276)
2 ASN00086077  145. -38.0  12.1 moora~  94870 <tibble>   (145.0964 -37.98)
3 ASN00086282  145. -37.7 113.  melbo~  94866 <tibble> (144.8321 -37.6655)
\end{CodeOutput}
\end{CodeChunk}

\hypertarget{tidyverse}{%
\subsection{Comparison to other spatio-temporal classes}\label{tidyverse}}

In \proglang{R}, there are other existing spatio-temporal data structures and this section compares and contrasts \pkg{cubble} with other existing alternatives, specifically \pkg{stars} and \pkg{sftime}. The \pkg{stars} package \citep{stars} uses an array structure, as opposed to a tibble, to represent multivariate spatio-temporal data. While both \pkg{stars} and \pkg{cubble} support vector and raster data, it is a matter of choice which structure to use given the application. Analysts working on satellite imageries may prefer the array structure in \pkg{stars}, while others originally working with spatio-temporal data in 2D data frames may find \pkg{cubble} easier to adopt from their existing computing workflow.

The \pkg{sftime} package \citep{sftime} also builds from a tibble object and its focus is on handling irregular spatio-temporal data. This means \pkg{sftime} can also handle full space-time grids and sparse space-time layouts represented in \pkg{cubble}. However, \pkg{cubble} uses nesting to avoid storing spatial variables repetitively at each timestamp. This provides memory efficiency when data is observed frequently, i.e.~daily or sub-daily, or the spatial geometry is computationally expensive to store repeatedly, i.e.~polygons or multipolygons. Consider the \code{climate_aus} data in the \pkg{cubble} package with 639 stations observed daily throughout the year 2020. In that case, the \code{sftime} object is approximately 14 times larger than the corresponding \code{cubble} object (118 MB vs.~8.5 MB).

\hypertarget{others}{%
\section{Other features and considerations}\label{others}}

\hypertarget{matching}{%
\subsection{Spatial and temporal matching}\label{matching}}

A useful task in spatio-temporal data analysis is to combine related temporal series within a close geographic neighborhood. For example, we may want to examine data from weather stations with water flow records from nearby river sensors to understand how precipitation relates to river levels. This might be useful for predicting potential for droughts and floods.

Matching temporal data across different locations from different data sources could be done by initially identifying the corresponding spatial locations between the two data sets. Subsequently, a set of temporal features can be calculated for the series at the selected locations that can be used to match the selected time series across locations. In cubble, locations from two datasets can be matched using the function \code{match_spatial()}. The function calculates the distance matrix of the locations between the two data sets and returns groups (\code{spatial_n_group}) with the smallest distances. For a given group, it is possible to include more locations with the argument \code{spatial_n_each} (default to 1 for one-on-one matching).

For the temporal matching a similarity score between the time series of spatially matched pairs is computed using the function \code{match_temporal()}. The similarity score is computed by a matching function which can be customized to any desired time series feature. The function \code{match_temporal()} takes as argument two time series in the form of a list and returns a single numerical value. By default, cubble uses a simple peak matching algorithm (\code{match_peak}) to count the number of peaks in two time series that fall within a specified time window.

The temporal matching requires two identifiers: one for separating each spatially matched group: \code{match_id} and one for separating the two data sources: \code{data_id}. Matching between different variables can be specified using the \code{temporal_by} argument, similar to the \code{by} syntax from dplyr's \code{*_join}.

\begin{verbatim}
match_temporal(
  <obj_from_match_spatial>, 
  data_id = ... , match_id = ..., 
  temporal_by = c("..." = "...")
)
\end{verbatim}

\hypertarget{interactive-graphics}{%
\subsection{Interactive graphics}\label{interactive-graphics}}

The cubble workflow neatly allows building an interactive graphics pipeline (e.g., \citet{buja1988elements}; \citet{buja1996interactive}; \citet{sutherland2000orca}; \citet{xie2014reactive}; \citet{cheng2016enabling}), simplifying the data pre-processing and preparing the ingredients for linked plots. Specifically, the spatial and temporal cubble correspond to the spatial and temporal visualisation, such as a map or a time series plot, that can be linked using functionality in the \pkg{crosstalk} \citep{crosstalk} package.

Figure \ref{fig:illu-interactive} illustrates the linking mechanism between a map and multiple time series. When a user selects a location on the map as shown on panel (a), the corresponding site is highlighted. This selection activates a row in the spatial cubble, which is then connected to the temporal cubble, resulting in the selection of all observations with the same ID as depicted in panel (b). Consequently the temporal cubble highlights the corresponding series in the time series plot displayed in panel (c). The linking can also be initiated from the time series plot by selecting points on the time series graph. This action selects rows with the same ID in the temporal cubble and the corresponding row in the spatial cubble so that points can be highlight on the map.

\begin{CodeChunk}
\begin{figure}

{\centering \includegraphics[width=1\linewidth,height=0.35\textheight]{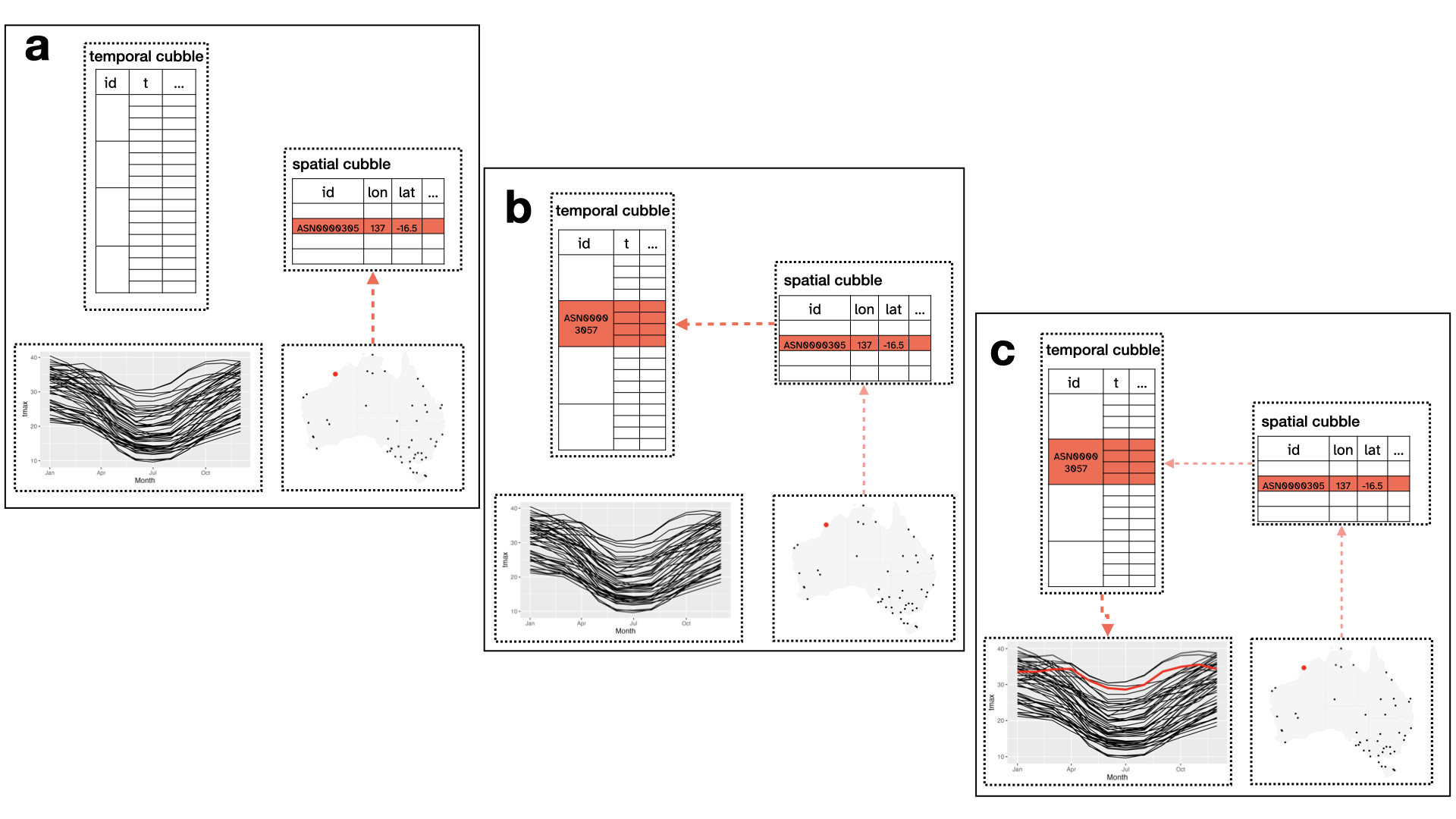} 

}

\caption{Linking between multiple plots is made possible by shared \code{crosstalk} objects. When a station is selected on the map (a), the corresponding row in the spatial \code{cubble} will be activated. This will activate the row with the same id in the temporal \code{cubble} (b) to trigger an update of the line plot (c). The cubble package makes linking between spatial and temporal plots easy.}\label{fig:illu-interactive}
\end{figure}
\end{CodeChunk}

\hypertarget{st_transformation}{%
\subsection{Spatio-temporal transformations}\label{st_transformation}}

Visualizing both space and time is important for exploring and understanding the data more completely, aiding in decision-making, and facilitating effective communication. Several approaches are common: facet maps across time, map animations, or interactive graphics that link maps and time series plots among others. Faceted maps and spatio-temporal animations focus on the spatial pattern making it difficult to assess temporal trends. The glyph map \citep{Wickham2012-yr} addresses this issue by plotting time series onto the map as a glyph.

It is achieved through a coordinate transformation. The transformation uses linear algebra to convert the temporal coordinates (minor coordinates) into the spatial coordinates (major coordinates) and is implemented in the package \pkg{GGally} \citep{ggally}. The \pkg{cubble} package provides a new \code{ggproto} implementation to create glyph maps, \code{geom_glyph()} requiring four aesthetics : \code{x_major}, \code{y_major}, \code{x_minor}, and \code{y_minor}:

\begin{verbatim}
data |> 
  ggplot() +
  geom_glyph(aes(x_major = ..., x_minor = ..., 
                 y_major = ..., y_minor = ...))
\end{verbatim}

Other useful controls to modify the glyph map that can be include are:

\begin{itemize}
\tightlist
\item
  the implementation of a polar coordinate glyph maps with \code{polar = TRUE},
\item
  the adjustment of the glyph size arguments using \code{width} and \code{height},
\item
  a transformation relative to the all series (\code{global_rescale} defaults to \code{TRUE}) or each single series, and
\item
  the XX of the reference boxes and lines with \code{geom_glyph_box()} and \code{geom_glyph_line()}.
\end{itemize}

\hypertarget{examples}{%
\section{Applications}\label{examples}}

Five examples are chosen to illustrate different aspects of the \pkg{cubble} package: (1) creating a \code{cubble} object from two Coronavirus (COVID) data tables with the challenge of having different location names, (2) using spatial transformations to make a glyph map of seasonal temperature changes, (3) matching river level data with weather station records to analyze water supply, (4) reading NetCDF format data to replicate a climate reanalysis plot, and (5) demonstrating the workflow to create interactively linked plots.

\hypertarget{covid}{%
\subsection{Victoria COVID spatio-temporal incidence and spread}\label{covid}}

Since the start of the COVID-19 pandemic, the Victoria State Government in Australia has been providing daily COVID-19 case counts per local government area (LGA). This data can be combined with map polygon data, available from the Australian Bureau of Statistics (ABS), to visualize COVID-19 incidence and spread. The COVID-19 count data (\code{covid}) and the LGA information (\code{lga}) are available in the \pkg{cubble} package as a \code{tsibble} object and an \code{sf} object, respectively. As is common, the different agencies have some difference in text id's labelling the spatial regions, so this example illustrates how this can be caught with \pkg{cubble}, and fixed. Discrepancies are flagged when creating the cubble object (see warnings in the R output below), notifying analysts of something that needs checking.

The \code{by} argument of the function \code{make_cubble()} is used to specify the spatial identifier in the two data sets:

\begin{CodeChunk}
\begin{CodeInput}
R> cb <- make_cubble(lga, covid, by = c("lga_name_2018" = "lga"))
\end{CodeInput}
\begin{CodeOutput}
Warning: st_centroid assumes attributes are constant over geometries
\end{CodeOutput}
\begin{CodeOutput}
! Some sites in the spatial table don't have temporal information
\end{CodeOutput}
\begin{CodeOutput}
! Some sites in the temporal table don't have spatial information
\end{CodeOutput}
\begin{CodeOutput}
! Use `check_key()` to check on the unmatched key
The cubble is created only with sites having both spatial and
temporal information
\end{CodeOutput}
\end{CodeChunk}

The difference in LGA naming between both data sets triggers a warning, alerting the user to this discrepancy. The warning message suggests there are some differences between the LGA encoding used by Victoria government and ABS. The mismatches can be checked using \code{check_key()}, which takes the same inputs as \code{make_cubble()}, but returns a summary of key matches between the spatial and temporal input data:

\begin{CodeChunk}
\begin{CodeInput}
R> (check_res <- check_key(
+   spatial = lga, temporal = covid, 
+   by = c("lga_name_2018" = "lga")
+ ))
\end{CodeInput}
\begin{CodeOutput}
$paired
# A tibble: 78 x 2
  spatial        temporal      
  <chr>          <chr>         
1 Alpine (S)     Alpine (S)    
2 Ararat (RC)    Ararat (RC)   
3 Ballarat (C)   Ballarat (C)  
4 Banyule (C)    Banyule (C)   
5 Bass Coast (S) Bass Coast (S)
# i 73 more rows

$potential_pairs
# A tibble: 2 x 2
  spatial             temporal    
  <chr>               <chr>       
1 Kingston (C) (Vic.) Kingston (C)
2 Latrobe (C) (Vic.)  Latrobe (C) 

$others
$others$spatial
character(0)

$others$temporal
[1] "Interstate" "Overseas"   "Unknown"   


attr(,"class")
[1] "key_tbl" "list"   
\end{CodeOutput}
\end{CodeChunk}

The result of the \code{check_key()} function is a list containing three elements: 1) matched keys from both tables, 2) potentially paired keys, and 3) others keys that can't be matched. Here, the main mismatch arises from the two LGAs: Kingston and Latrobe (Kingston is an LGA in both Victoria and South Australia and Latrobe is an LGA in both Victoria and Tasmania). Analysts can then reconcile the spatial and temporal data based on this summary and recreate the cubble object:

\begin{CodeChunk}
\begin{CodeInput}
R> lga2 <- lga |>
+   rename(lga = lga_name_2018) |> 
+   mutate(lga = ifelse(lga == "Kingston (C) (Vic.)", "Kingston (C)", lga),
+          lga = ifelse(lga == "Latrobe (C) (Vic.)", "Latrobe (C)", lga))
R>   
R> covid2 <- covid |> filter(!lga %in% check_res$others$temporal)
R> 
R> (cb <- make_cubble(spatial = lga2, temporal = covid2))
\end{CodeInput}
\begin{CodeOutput}
# cubble:   key: lga [80], index: date, nested form, [sf]
# spatial:  [140.961682, -39.1339581, 149.976291, -33.9960517], WGS 84
# temporal: date [date], n [dbl], avg_7day [dbl]
  lga             long   lat                                   geometry ts      
  <chr>          <dbl> <dbl>                             <GEOMETRY [°]> <list>  
1 Alpine (S)      147. -36.9 POLYGON ((146.7258 -36.45922, 146.7198 -3~ <tbl_ts>
2 Ararat (RC)     143. -37.5 POLYGON ((143.1807 -37.73152, 143.0609 -3~ <tbl_ts>
3 Ballarat (C)    144. -37.5 POLYGON ((143.6622 -37.57241, 143.68 -37.~ <tbl_ts>
4 Banyule (C)     145. -37.7 POLYGON ((145.1357 -37.74091, 145.1437 -3~ <tbl_ts>
5 Bass Coast (S)  146. -38.5 MULTIPOLYGON (((145.5207 -38.30667, 145.5~ <tbl_ts>
# i 75 more rows
\end{CodeOutput}
\end{CodeChunk}

\hypertarget{historicaltmax}{%
\subsection{Australian historical maximum temperature}\label{historicaltmax}}

The Global Historical Climatology Network (GHCN) provides daily climate measures for stations worldwide. In the \pkg{cubble} package, the cubble object \code{historical_tmax} contains daily maximum temperature data for 75 stations in Australia, covering two periods: 1971-1975 and 2016-2020. This example uses \pkg{dplyr} verbs to wrangle a cubble object, and pivot between the spatial and temporal form for different parts of the analysis. The result is glyph maps to compare the changes in temperature between these two periods, created with \pkg{ggplot2}.

To prevent overlapping of weather stations on the map, stations are selected to ensure a minimum distance of 50km. Distance between stations can be calculated with \code{sf::st_distance()} after turning the spatial cubble to also be an sf object with \code{make_spatial_sf()}:

\begin{CodeChunk}
\begin{CodeInput}
R> a <- historical_tmax |> make_spatial_sf() |> st_distance()
R> a[upper.tri(a, diag = TRUE)] <- 1e6
R> 
R> (tmax <- historical_tmax |> 
+   filter(rowSums(a < units::as_units(50, "km")) == 0))
\end{CodeInput}
\begin{CodeOutput}
# cubble:   key: id [54], index: date, nested form
# spatial:  [141.2652, -39.1297, 153.3633, -28.9786], Missing CRS!
# temporal: date [date], tmax [dbl]
  id           long   lat  elev name                     wmo_id ts      
  <chr>       <dbl> <dbl> <dbl> <chr>                     <dbl> <list>  
1 ASN00047016  141. -34.0    43 lake victoria storage     94692 <tibble>
2 ASN00047019  142. -32.4    61 menindee post office      94694 <tibble>
3 ASN00048015  147. -30.0   115 brewarrina hospital       95512 <tibble>
4 ASN00048027  146. -31.5   260 cobar mo                  94711 <tibble>
5 ASN00048031  149. -29.5   145 collarenebri (albert st)  95520 <tibble>
# i 49 more rows
\end{CodeOutput}
\end{CodeChunk}

The daily maximum temperature is then averaged into monthly series for each period within the temporal cube. In the code above, the last step with \code{unfold()} moves the two coordinate columns (\code{long, lat}) into the temporal cubble, preparing the data for the construction of a glyph map:

\begin{CodeChunk}
\begin{CodeInput}
R> (tmax <- tmax |>
+   face_temporal() |> 
+   group_by(
+     yearmonth = tsibble::make_yearmonth(
+       year = ifelse(lubridate::year(date) > 2015, 2016, 1971),
+       month = lubridate::month(date))
+   )|>
+   summarise(tmax = mean(tmax, na.rm = TRUE)) |> 
+   mutate(group = as.factor(lubridate::year(yearmonth)),
+          month = lubridate::month(yearmonth)) |> 
+   unfold(long, lat))
\end{CodeInput}
\begin{CodeOutput}
# cubble:   key: id [54], index: yearmonth, long form
# temporal: 1971 Jan -- 2016 Dec [1M], has gaps!
# spatial:  long [dbl], lat [dbl], elev [dbl], name [chr], wmo_id [dbl]
  yearmonth id           tmax group month  long   lat
      <mth> <chr>       <dbl> <fct> <dbl> <dbl> <dbl>
1  1971 Jan ASN00047016  31.1 1971      1  141. -34.0
2  1971 Jan ASN00047019  33.1 1971      1  142. -32.4
3  1971 Jan ASN00048015  33.9 1971      1  147. -30.0
4  1971 Jan ASN00048027  32.5 1971      1  146. -31.5
5  1971 Jan ASN00048031  33.3 1971      1  149. -29.5
# i 1,276 more rows
\end{CodeOutput}
\end{CodeChunk}

The code below counts the number of observations for each location, revealing that there are several with less than 24 observations -- these stations lack temperature values for some months. In this example, those stations are removed by switching to the spatial cubble to operate on the spatial component over time, and then, move back into the temporal cubble (to make the glyph map):

\begin{CodeChunk}
\begin{CodeInput}
R> tmax <- tmax |> 
+   face_spatial() |> 
+   rowwise() |>
+   filter(nrow(ts) == 24) |>
+   face_temporal()
\end{CodeInput}
\end{CodeChunk}

The following code creates the glyph map (a) in Figure \ref{fig:glyphmap} (additional codes are needed for highlighting the single station, Cobar and styling) and the glyph map (c) is produced similarly after further processing the data.

\begin{verbatim}
nsw_vic <- ozmaps::abs_ste |> 
  filter(NAME %in% c("Victoria","New South Wales"))

tmax |> 
  ggplot(aes(x_major = long, x_minor = month, 
             y_major = lat, y_minor = tmax,
             group = interaction(id, group))) + 
  geom_sf(data =  nsw_vic, ...,  inherit.aes = FALSE) + 
  geom_glyph_box(width = 0.8, height = 0.3) + 
  geom_glyph(aes(color = group), width = 0.8, height = 0.3) +
  ...
\end{verbatim}

\begin{CodeChunk}
\begin{figure}

{\centering \includegraphics[width=1\linewidth]{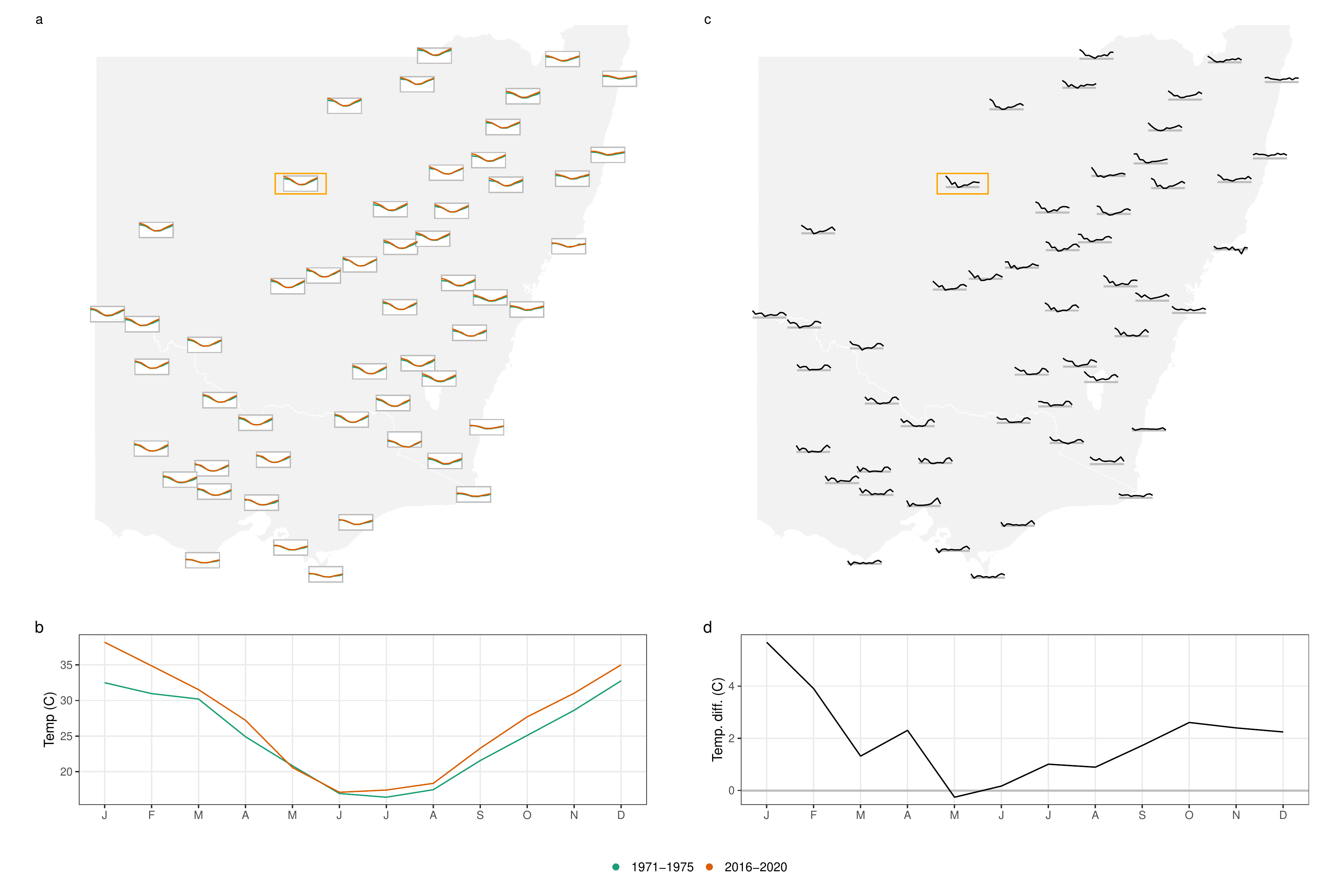} 

}

\caption[Glyph maps comparing temperature change between 1971-1975 and 2016-2020 for 54 stations in Victoria and New South Wales, Australia]{Glyph maps comparing temperature change between 1971-1975 and 2016-2020 for 54 stations in Victoria and New South Wales, Australia. Overlaid line plots show monthly temperature (a) where a hint of late summer warming can be seen. Transforming to temperature differences (c) shows pronounced changes between the two periods. The horizontal guideline marks zero difference. One station, Cobar, is highlighted in the glyph maps and shown separately (b, d). Here the late summer (Jan-Feb) warming pattern, which is more prevalent at inland locations, is clear.}\label{fig:glyphmap}
\end{figure}
\end{CodeChunk}

\hypertarget{river-levels-and-rainfall-in-victoria}{%
\subsection{River levels and rainfall in Victoria}\label{river-levels-and-rainfall-in-victoria}}

This example illustrates spatio-temporal matching of water level data from stream sensors with precipitation data from climate weather stations. The data \code{river} from the \pkg{cubble} package contains water course level data for 71 river gauges collected in Victoria, Australia. Victoria weather station data can be extracted from the \code{climate_aus} data in the \pkg{cubble} package. This example demonstrates the use of the matching function introduced in Section \ref{matching} to find river gauges that mirror changes in precipitation regimes captured in the climate weather stations in Victoria.

\begin{CodeChunk}
\begin{CodeInput}
R> climate_vic <- climate_aus |>
+   filter(between(as.numeric(substr(id, 7, 8)), 76, 90)) |>
+   mutate(type = "climate")
R> river <- cubble::river |> mutate(type = "river") 
\end{CodeInput}
\end{CodeChunk}

We assume that matches would be close spatially and have a similar temporal pattern, accounting for scale (volume of rainfall relative to volume of stream capacity) and temporal lags (time to take water to travel to location). This motivates a two-step approach: spatial match is first performed between both data sets, followed by a temporal pattern match on a reduced set of locations.

With \code{match_spatial()}, we can obtain a summary of the closest pairs (here it is 10) of weather stations and river gauges to print or as a list of matched cubbles (if \code{return_cubble = TRUE}). The following shows the print results:

\begin{CodeChunk}
\begin{CodeInput}
R> res_sp <- match_spatial(df1 = climate_vic, df2 = river, 
+                         spatial_n_group = 10)
R> print(res_sp)
\end{CodeInput}
\begin{CodeOutput}
# A tibble: 10 x 4
  from        to      dist group
  <chr>       <chr>    [m] <int>
1 ASN00088051 406213 1838.     1
2 ASN00084145 222201 2185.     2
3 ASN00085072 226027 3282.     3
4 ASN00080015 406704 4034.     4
5 ASN00085298 226027 4207.     5
# i 5 more rows
\end{CodeOutput}
\end{CodeChunk}

Notice from the printed list that river station 226027 matches two weather stations. This is possible, and is why printing the matching results first is preferable. It is possible to keep multiple matches in order to find the best temporal match among the set using the \code{spatial_n_each} argument. In combination with \code{spatial_n_group} more complicated matched pairs can be provided to the temporal matching. For example, if \code{spatial_n_each=4} and \code{spatial_n_group=2}, two groups each with the closest four spatial neighbors would be created.

The argument \code{return_cubble = TRUE} can be used to return the match as a list of matched cubbles. This list is of length \code{saptial_n_group} and each cubble element has 2x\code{saptial_n_each} rows. The subsequent temporal matches can be then performed on each cubble element using the \code{purrr::map()} or \code{lapply()} syntax.

In the case of matched pairs (\code{spatial_n_each = 1} as default here), the list can be converted to a single cubble with the \code{bind_rows()} function. The matched pair/group \(i\) will be stored in row \(2i-1\) and \(2i\) in the augmented cubble. However, the user would need to decide how to handle the multiple matches, as observed in this example, for example by removing one or aggregating results, because a cubble requires unique IDs. This example creates the list of matched cubbles after excluding the two pairs where a river station is matched to more than one weather station (river station 226027 is matched twice in group 3 and 5 and similarly for station 230200 in group 7 and 8). The temporal matching would follow on each pair/group \(i\).

\begin{CodeChunk}
\begin{CodeInput}
R> res_sp <- match_spatial(
+   df1 = climate_vic, df2 = river, 
+   spatial_n_group = 10, return_cubble = TRUE)
R> (res_sp <- res_sp[-c(5, 8)] |> bind_rows())
\end{CodeInput}
\begin{CodeOutput}
# cubble:   key: id [16], index: date, nested form, [sf]
# spatial:  [144.5203, -38.144913, 148.4667, -36.128657], WGS 84
# temporal: date [date], prcp [dbl], tmax [dbl], tmin [dbl]
  id      long   lat  elev name  wmo_id ts       type              geometry
  <chr>  <dbl> <dbl> <dbl> <chr>  <dbl> <list>   <chr>          <POINT [°]>
1 ASN00~  145. -37.0 290   rede~  94859 <tibble> clim~  (144.5203 -37.0194)
2 406213  145. -37.0  NA   CAMP~     NA <tibble> river (144.5403 -37.01512)
3 ASN00~  148. -37.7  62.7 orbo~  95918 <tibble> clim~  (148.4667 -37.6922)
4 222201  148. -37.7  NA   SNOW~     NA <tibble> river  (148.451 -37.70739)
5 ASN00~  147. -38.1   4.6 east~  94907 <tibble> clim~  (147.1322 -38.1156)
# i 11 more rows
# i 2 more variables: group <int>, dist [m]
\end{CodeOutput}
\end{CodeChunk}

To match the water level and precipitation time series across the matched locations, the function \code{match_temporal()} is used with the variables \code{group} and \code{type} identifying the matching group \(i\), and the two data sources:

\begin{CodeChunk}
\begin{CodeInput}
R> (res_tm <- match_temporal(data = res_sp,
+                           data_id = type, match_id = group,
+                           temporal_by = c("prcp" = "Water_course_level")))
\end{CodeInput}
\begin{CodeOutput}
# A tibble: 8 x 2
  group match_res
  <int>     <dbl>
1     1        30
2     2         5
3     3        14
4     4        20
5     6        23
# i 3 more rows
\end{CodeOutput}
\end{CodeChunk}

Similarly, the cubble output can be returned using the argument \code{return_cubble = TRUE}. Here, we select the four pairs of time series (precipitation/water level) with the highest number of matching peaks and show them on the map (Figure \ref{fig:matching} a). The time series of river levels is standardized to make the comparison easier in panel (b).

\begin{CodeChunk}
\begin{CodeInput}
R> res_tm <- match_temporal(data =  res_sp,
+                          data_id = type, match_id = group,
+                          temporal_by = c("prcp" = "Water_course_level"),
+                          return_cubble = TRUE)
R> (res_tm <- res_tm |> bind_rows() |> filter(group %in% c(1, 7, 6, 9)))
\end{CodeInput}
\begin{CodeOutput}
# cubble:   key: id [8], index: date, nested form, [sf]
# spatial:  [144.5203, -37.8817, 147.572223, -36.8472], WGS 84
# temporal: date [date], matched [dbl]
  id         long   lat  elev name  wmo_id type              geometry group
  <chr>     <dbl> <dbl> <dbl> <chr>  <dbl> <chr>          <POINT [°]> <int>
1 ASN00088~  145. -37.0 290   rede~  94859 clim~  (144.5203 -37.0194)     1
2 406213     145. -37.0  NA   CAMP~     NA river (144.5403 -37.01512)     1
3 ASN00082~  146. -36.8 502   stra~  95843 clim~  (145.7308 -36.8472)     6
4 405234     146. -36.9  NA   SEVE~     NA river (145.6828 -36.88701)     6
5 ASN00086~  145. -37.7  78.4 esse~  95866 clim~  (144.9066 -37.7276)     7
# i 3 more rows
# i 3 more variables: dist [m], ts <list>, match_res <dbl>
\end{CodeOutput}
\end{CodeChunk}

\begin{CodeChunk}
\begin{figure}

{\centering \includegraphics[width=1\linewidth]{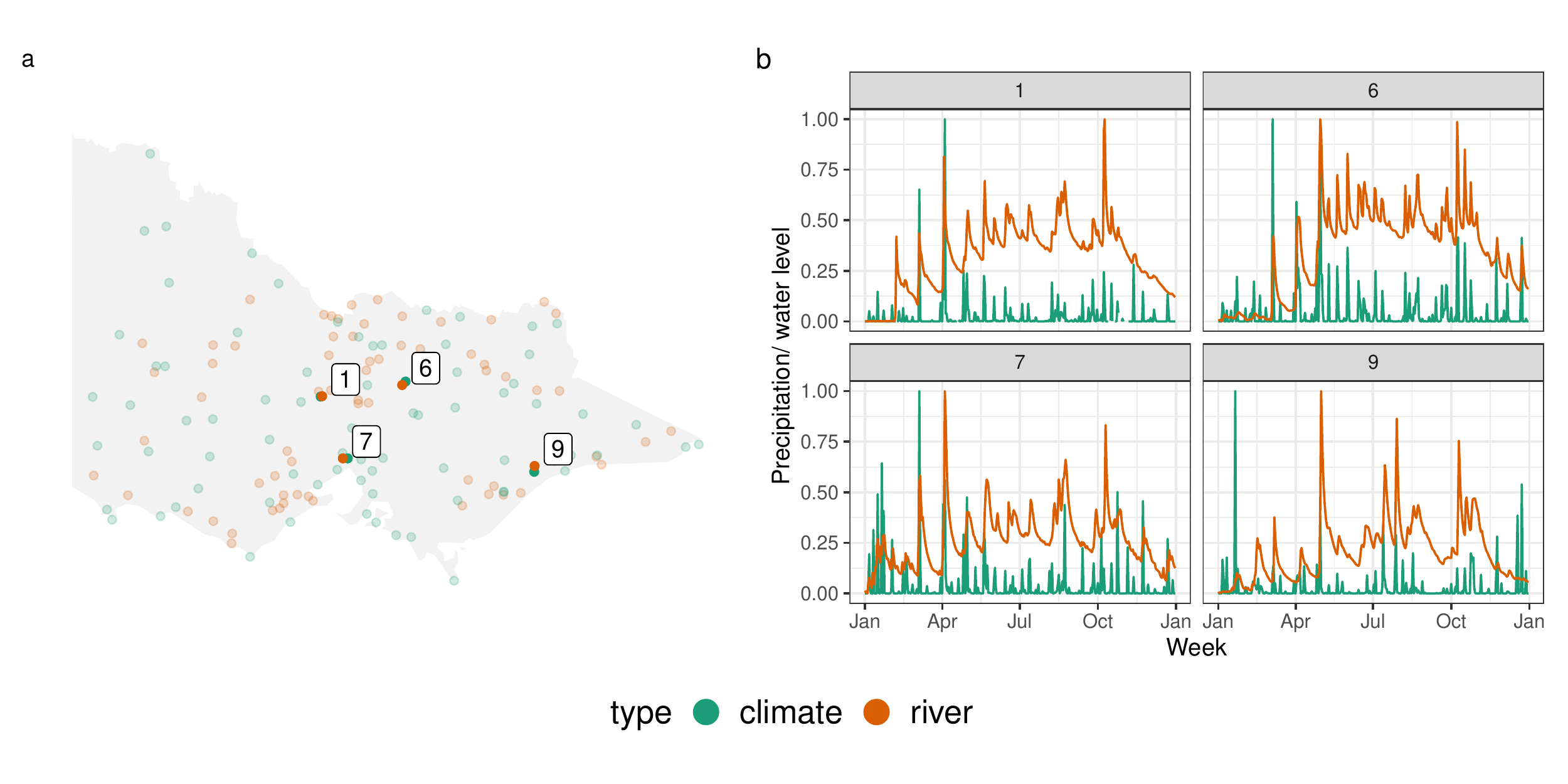} 

}

\caption[Example of matching weather stations and river gauges]{Example of matching weather stations and river gauges. These four station pairs shown on the map (a) and as time series plots (b) would be considered to be matching. Precipitation and water level have been standardised between 0 and 1 to be displayed in the same scale in (b). The peaks in the time series roughly match, and would reflect precipitation increasing water levels.}\label{fig:matching}
\end{figure}
\end{CodeChunk}

\hypertarget{era5-climate-reanalysis-data}{%
\subsection{ERA5: climate reanalysis data}\label{era5-climate-reanalysis-data}}

The ERA5 reanalysis \citep{hersbach2020era5} provides hourly estimates of atmospheric, land and oceanic climate variables on a global scale and is available in the NetCDF format from Copernicus Climate Data Store (CDS). This example demonstrates a case of analysing raster spatio-temporal data using cubble, replicating Figure 19 from the \citet{hersbach2020era5} paper. The plot shows the southern polar vortex splitting into two on 2002-09-26, and further splitting into four on 2002-10-04. Further explanation of why this is interesting can be found in the figure source, and also in \citet{simmons2020global} and \citet{simmons2005ecmwf}.

A \code{ncdf4} object \citep{ncdf4} can be converted into a cubble using \code{as_cubble()} and the NetCDF data can be subsetted with arguments \code{vars}, \code{long_range} and \code{lat_range}. In this example, the variables q (specific humidity) and z (geopotential) are read in and the coordinates are subsetted to every degree in longitude and latitude:

\begin{CodeChunk}
\begin{CodeInput}
R> raw <- ncdf4::nc_open(here::here("data/era5-pressure.nc"))
R> (dt <- as_cubble(
+   raw, vars = c("q", "z"),
+   long_range = seq(-180, 180, 1), lat_range = seq(-88, -15, 1)))
\end{CodeInput}
\begin{CodeOutput}
# cubble:   key: id [26640], index: time, nested form
# spatial:  [-180, -88, 179, -15], Missing CRS!
# temporal: time [date], q [dbl], z [dbl]
     id  long   lat ts              
  <int> <dbl> <dbl> <list>          
1     1  -180   -15 <tibble [8 x 3]>
2     2  -179   -15 <tibble [8 x 3]>
3     3  -178   -15 <tibble [8 x 3]>
4     4  -177   -15 <tibble [8 x 3]>
5     5  -176   -15 <tibble [8 x 3]>
# i 26,635 more rows
\end{CodeOutput}
\end{CodeChunk}

Once the NetCDF data is coerced into a cubble object, subsequent analysis can be conducted to filter on the date of interest, scale the variable specific humidity and create visualisation in ggplot to reproduce the ERA5 plot. A snippet of code to create Figure \ref{fig:netcdf} is provided below with additional codes needed to style the plot.

\begin{verbatim}
res <- dt |> 
  face_temporal() |> 
  filter(lubridate::date(time) %in% 
           as.Date(c("2002-09-22", "2002-09-26",
                     "2002-09-30", "2002-10-04"))) |>
  unfold(long, lat) |> 
  mutate(q = q* 10^6)

con <- rnaturalearth::ne_coastline("small", returnclass = "sf")
box <- st_bbox(c(xmin = -180, ymin = -90, xmax = 180, ymax = -15), 
               crs = st_crs(con)) 
country <- con |> 
  st_geometry() |> 
  st_crop(box) |> 
  st_cast("MULTILINESTRING")

res |> 
  ggplot() +
  geom_point(aes(x = long, y = lat, color = q)) + 
  geom_contour(data = res, aes(x = long, y = lat, z = z), ...) +
  geom_sf(data = country, ...) +
  ...
\end{verbatim}

\begin{CodeChunk}
\begin{figure}

{\centering \includegraphics[width=1\linewidth]{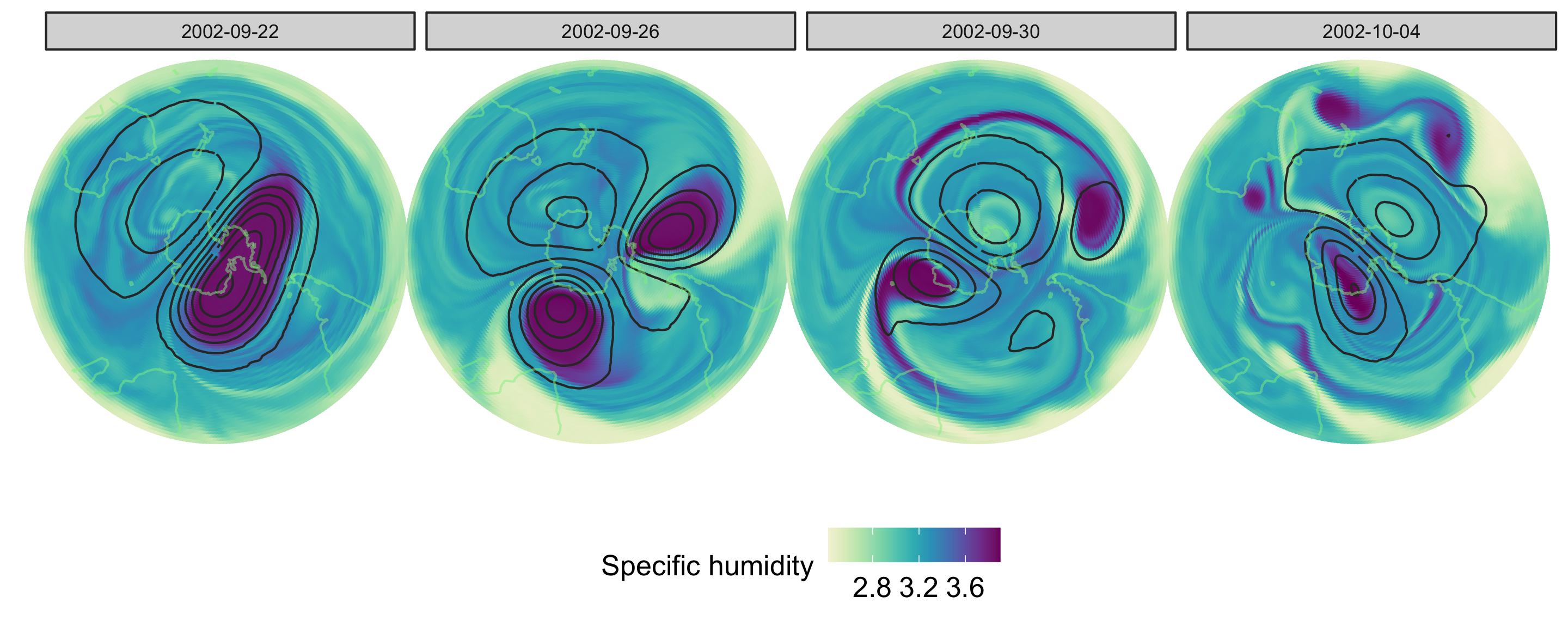} 

}

\caption[An example illustrating that cubble can be used to readily reproduce common spatiotemporal analyses]{An example illustrating that cubble can be used to readily reproduce common spatiotemporal analyses. This plot of ERA5 reanalysis (Fig. 19, Hersbach et al, 2020) shows the break-up of the southern polar vortex in late September and early October 2002. The polar vortex, signalled by the high specific humidity, splits into two on 2002-09-26 and further splits into four on 2002-10-04.}\label{fig:netcdf}
\end{figure}
\end{CodeChunk}

\hypertarget{australian-temperature-range}{%
\subsection{Australian temperature range}\label{australian-temperature-range}}

Interactive graphics can be especially useful for spatio-temporal data because they make it possible to look at the data in multiple ways on-the-fly. This last example describes the process of using
\pkg{cubble} with the \pkg{crosstalk} package to build an interactive display connecting a map of Australia, with ribbon plots of temperature range observed at a group of stations in 2020. The purpose is to explore the variation of monthly temperature range over the country.

Firstly, we summarise the daily data in \code{climate_aus} into monthly averages and calculate the variance of the monthly averages differences between the minimum and maximum temperatures. This variance will be used to color the temperature band later.

\begin{verbatim}
clean <- climate_aus |>
  face_temporal() |> 
  mutate(month = lubridate::month(date)) |>
  group_by(month) |>
  summarise(
    tmax = mean(tmax, na.rm = TRUE),
    tmin = mean(tmin, na.rm = TRUE),
    diff = mean(tmax - tmin, na.rm = TRUE)
  ) |> 
  face_spatial() |> 
  rowwise() |>
  mutate(temp_diff_var = var(ts$diff, na.rm =TRUE))
\end{verbatim}

The spatial and temporal cubble are then created into shared \code{crosstalk} objects, plotted as ggplots, and combined together using \code{crosstalk::bscols()}:

\begin{verbatim}
sd_spatial <- clean |> SharedData$new(~id, group = "cubble")

sd_temporal <- clean |> 
  face_temporal() |> 
  SharedData$new(~id, group = "cubble")
  
p1 <- sd_spatial |> ggplot() + ...
p2 <- sd_temporal |> ggplot() + ...
crosstalk::bscols(plotly::ggplotly(p1), plotly::ggplotly(p2), ...)
\end{verbatim}

Figure \ref{fig:interactive-linking} shows three snapshots of the interactivity. Plot (a) shows the initial state of the interactive display: all locations are shown as dots on the map, coloured by the temperature range, and the right plot shows the ribbons representing maximum to minimum for all stations. In plot (b) the station shows a high variance on the initial map, the ``Mount Elizabeth'' station, is selected and this produces the ribbon on the right. In plot (c) the lowest temperature in August is selected on the left map and this corresponds to the ``Thredbo'' station in the mountain area in Victoria and New South Wales. This station is compared to a station in the Tasmania island, the southernmost island of the country, selected on the map.

\begin{CodeChunk}
\begin{figure}

{\centering \includegraphics[width=1\linewidth,height=0.23\textheight]{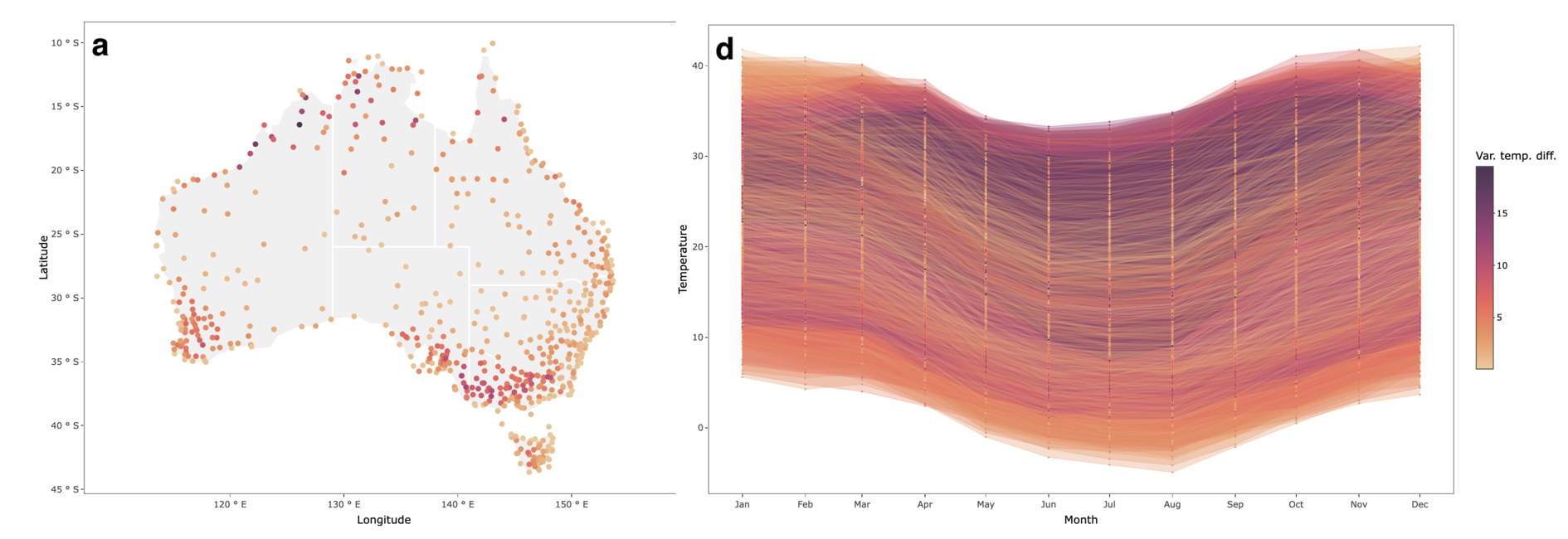} \includegraphics[width=1\linewidth,height=0.23\textheight]{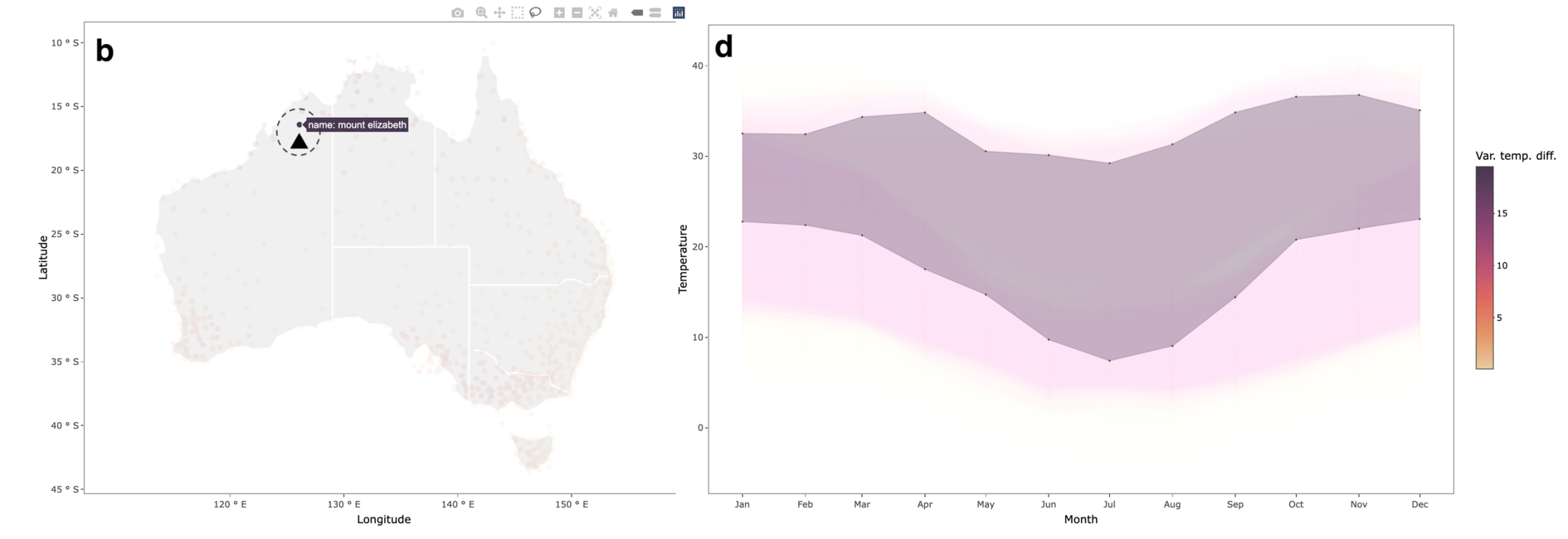} \includegraphics[width=1\linewidth,height=0.23\textheight]{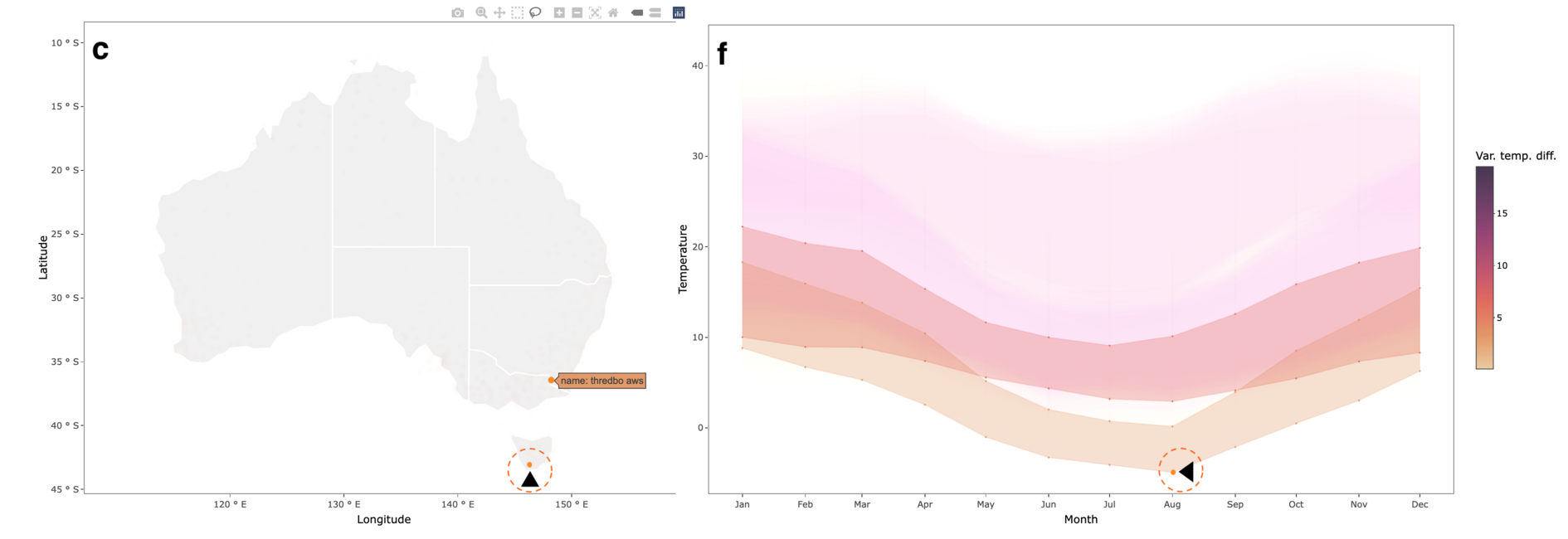} 

}

\caption[Illustration of using cubble for interactive graphics]{Illustration of using cubble for interactive graphics. Here we explore temperature variation by linking a map and a seasonal display. Each row is a screen dump of the process. The top row shows all locations and all temperature profiles. Selecting a particular location on the map (here Mount Elizabeth) produces the plot in the second row. The maximum and minimum temperatures are shown using a ribbon. The bottom row first selects the lowest temperature in August in the seasonal display from the first row, which highlights the corresponding station on the map (Thredbo). Another  station, located in the Tasmania Island, is then selected to compare its temperature variation with the Thredbo station.}\label{fig:interactive-linking}
\end{figure}
\end{CodeChunk}

\hypertarget{conclude}{%
\section{Conclusion}\label{conclude}}

This paper presents the \proglang{R} package \pkg{cubble} for organizing, wrangling and visualizing spatio-temporal data. The package introduces a new data structure, \code{cubble}, consisting of two subclasses, spatial cubble and a temporal cubble, to organise spatio-temporal data in two different formats within the tidy data framework. The data structure and functions introduced in the package can be used and combined with existing tools for data wrangling, spatial and temporal data analysis, and visualization.

The paper includes several examples to illustrate how \pkg{cubble} is useful for spatio-temporal analysis. These examples cover different tasks of a typical data analysis workflow: handling data with spatial and temporal misalignment, matching data from multiple sources, and creating both static and interactive graphics. In addition, a re-working of an existing climate reanalysis using cubble is explained.

Possible future improvements would be in two main directions: handling much larger data, and integrating smoothly with modeling. Our work has been designed to make it easier to focus on intricate space-time patterns, and it expects that the analyst first reduces the data if the spatio-temporal data is huge, before creating a cubble. Additional tools for pre-processing large data would be helpful. Because the cubble structure is based on tidy data principles, it potentially integrates well with tidy tools for models, as provided by the \pkg{tidymodels} project. Although that project focuses on providing an interface to the vast array of statistical and machine learning models, it would be useful to extend its capacity to provide a more unified interface to many spatio-temporal modeling software.

\hypertarget{acknowledgement}{%
\section{Acknowledgement}\label{acknowledgement}}

This work is funded by a Commonwealth Scientific and Industrial Research Organisation (CSIRO) Data61 Scholarship and started while Nicolas Langrené was affiliated with CSIRO's Data61. The article is created using the package \pkg{knitr} \citep{knitr} and \pkg{rmarkdown} \citep{rmarkdown} in \proglang{R} with the \code{rticles::jss_article} template. The source code for reproducing this paper can be found at: \url{https://github.com/huizezhang-sherry/paper-cubble}.

\bibliography{references.bib}

\begin{thebibliography}{24}
\newcommand{\enquote}[1]{``#1''}
\providecommand{\natexlab}[1]{#1}
\providecommand{\url}[1]{\texttt{#1}}
\providecommand{\urlprefix}{URL }
\expandafter\ifx\csname urlstyle\endcsname\relax
  \providecommand{\doi}[1]{doi:\discretionary{}{}{}#1}\else
  \providecommand{\doi}{doi:\discretionary{}{}{}\begingroup \urlstyle{rm}\Url}\fi
\providecommand{\eprint}[2][]{\url{#2}}

\bibitem[{Bivand \emph{et~al.}(2008)Bivand, Pebesma, G{\'o}mez-Rubio, and Pebesma}]{bivand2008applied}
Bivand RS, Pebesma EJ, G{\'o}mez-Rubio V, Pebesma EJ (2008).
\newblock \emph{Applied spatial data analysis with R}, volume 747248717.
\newblock Springer.

\bibitem[{Buja \emph{et~al.}(1988)Buja, Asimov, and Hurley}]{buja1988elements}
Buja A, Asimov D, Hurley C (1988).
\newblock \enquote{Elements of A Viewing Pipeline.}
\newblock \emph{Dynamic Graphics Statistics}, p. 277.

\bibitem[{Buja \emph{et~al.}(1996)Buja, Cook, and Swayne}]{buja1996interactive}
Buja A, Cook D, Swayne DF (1996).
\newblock \enquote{Interactive High-dimensional Data Visualization.}
\newblock \emph{Journal of Computational and Graphical Statistics}, \textbf{5}(1), 78--99.
\newblock \urlprefix\url{https://doi.org/10.2307/1390754}.

\bibitem[{Cheng and Sievert(2021)}]{crosstalk}
Cheng J, Sievert C (2021).
\newblock \emph{\pkg{crosstalk}: Inter-Widget Interactivity for HTML Widgets}.
\newblock R package version 1.1.1, \urlprefix\url{https://CRAN.R-project.org/package=crosstalk}.

\bibitem[{Cheng \emph{et~al.}(2016)Cheng, Cook, and Hofmann}]{cheng2016enabling}
Cheng X, Cook D, Hofmann H (2016).
\newblock \enquote{Enabling Interactivity on Displays of Multivariate Time Series and Longitudinal Data.}
\newblock \emph{Journal of Computational and Graphical Statistics}, \textbf{25}(4), 1057--1076.
\newblock \urlprefix\url{https://doi.org/10.1080/10618600.2015.1105749}.

\bibitem[{Hersbach \emph{et~al.}(2020)Hersbach, Bell, Berrisford, Hirahara, Hor{á}nyi, Mu{ñ}oz-Sabater, Nicolas, Peubey, Radu, Schepers \emph{et~al.}}]{hersbach2020era5}
Hersbach H, Bell B, Berrisford P, Hirahara S, Hor{á}nyi A, Mu{ñ}oz-Sabater J, Nicolas J, Peubey C, Radu R, Schepers D, \emph{et~al.} (2020).
\newblock \enquote{The ERA5 Global Reanalysis.}
\newblock \emph{Quarterly Journal of the Royal Meteorological Society}, \textbf{146}(730), 1999--2049.

\bibitem[{Lovelace \emph{et~al.}(2019)Lovelace, Nowosad, and Muenchow}]{lovelace2019geocomputation}
Lovelace R, Nowosad J, Muenchow J (2019).
\newblock \emph{Geocomputation with R}.
\newblock CRC Press.

\bibitem[{Menne \emph{et~al.}(2012)Menne, Durre, Vose, Gleason, and Houston}]{menne2012overview}
Menne MJ, Durre I, Vose RS, Gleason BE, Houston TG (2012).
\newblock \enquote{An overview of the global historical climatology network-daily database.}
\newblock \emph{Journal of atmospheric and oceanic technology}, \textbf{29}(7), 897--910.

\bibitem[{Pebesma(2012)}]{spacetime}
Pebesma E (2012).
\newblock \enquote{\pkg{spacetime}: Spatio-Temporal Data in \proglang{R}.}
\newblock \emph{Journal of Statistical Software}, \textbf{51}(7), 1–30.
\newblock \urlprefix\url{https://doi.org/10.18637/jss.v051.i07}.

\bibitem[{Pebesma(2021)}]{stars}
Pebesma E (2021).
\newblock \emph{\pkg{stars}: Spatiotemporal Arrays, Raster and Vector Data Cubes}.
\newblock R package version 0.5-2, \urlprefix\url{https://CRAN.R-project.org/package=stars}.

\bibitem[{Pebesma and Bivand(2019)}]{pebesma2019spatial}
Pebesma E, Bivand R (2019).
\newblock \enquote{Spatial data science.}

\bibitem[{Pebesma and Bivand(2022)}]{ctvspatiotemporal}
Pebesma E, Bivand R (2022).
\newblock \enquote{{CRAN} Task View: Handling and Analyzing Spatio-Temporal Data.}
\newblock Version~2022-03-07, \urlprefix\url{https://CRAN.R-project.org/view=SpatioTemporal}.

\bibitem[{Pierce(2019)}]{ncdf4}
Pierce D (2019).
\newblock \emph{\pkg{ncdf4}: Interface to Unidata netCDF (Version 4 or Earlier) Format Data Files}.
\newblock R package version 1.17, \urlprefix\url{https://CRAN.R-project.org/package=ncdf4}.

\bibitem[{Schloerke \emph{et~al.}(2021)Schloerke, Cook, Larmarange, Briatte, Marbach, Thoen, Elberg, and Crowley}]{ggally}
Schloerke B, Cook D, Larmarange J, Briatte F, Marbach M, Thoen E, Elberg A, Crowley J (2021).
\newblock \emph{\pkg{GGally}: Extension to \pkg{ggplot2}}.
\newblock R package version 2.1.2, \urlprefix\url{https://CRAN.R-project.org/package=GGally}.

\bibitem[{Simmons \emph{et~al.}(2005)Simmons, Hortal, Kelly, McNally, Untch, and Uppala}]{simmons2005ecmwf}
Simmons A, Hortal M, Kelly G, McNally A, Untch A, Uppala S (2005).
\newblock \enquote{ECMWF Analyses and Forecasts of Stratospheric Winter Polar Vortex Breakup: September 2002 in the Southern Hemisphere and Related Events.}
\newblock \emph{Journal of the Atmospheric Sciences}, \textbf{62}(3), 668 -- 689.
\newblock \doi{10.1175/JAS-3322.1}.
\newblock \urlprefix\url{https://journals.ametsoc.org/view/journals/atsc/62/3/jas-3322.1.xml}.

\bibitem[{Simmons \emph{et~al.}(2020)Simmons, Soci, Nicolas, Bell, Berrisford, Dragani, Flemming, Haimberger, Healy, Hersbach, Hor{\'a}nyi, Inness, Munoz-Sabater, Radu, and Schepers}]{simmons2020global}
Simmons A, Soci C, Nicolas J, Bell B, Berrisford P, Dragani R, Flemming J, Haimberger L, Healy S, Hersbach H, Hor{\'a}nyi A, Inness A, Munoz-Sabater J, Radu R, Schepers D (2020).
\newblock \enquote{Global Stratospheric Temperature Bias and Other Stratospheric Aspects of ERA5 and ERA5.1.}
\newblock (859).
\newblock \doi{10.21957/rcxqfmg0}.
\newblock \urlprefix\url{https://www.ecmwf.int/node/19362}.

\bibitem[{Sutherland \emph{et~al.}(2000)Sutherland, Rossini, Lumley, Lewin-Koh, Dickerson, Cox, and Cook}]{sutherland2000orca}
Sutherland P, Rossini A, Lumley T, Lewin-Koh N, Dickerson J, Cox Z, Cook D (2000).
\newblock \enquote{\pkg{Orca}: A Visualization Toolkit for High-dimensional Data.}
\newblock \emph{Journal of Computational and Graphical Statistics}, \textbf{9}(3), 509--529.
\newblock \urlprefix\url{https://www.tandfonline.com/doi/abs/10.1080/10618600.2000.10474896}.

\bibitem[{Teickner \emph{et~al.}(2022)Teickner, Pebesma, and Graeler}]{sftime}
Teickner H, Pebesma E, Graeler B (2022).
\newblock \emph{\pkg{sftime}: Classes and Methods for Simple Feature Objects that Have a Time Column}.
\newblock Https://r-spatial.github.io/sftime/, https://github.com/r-spatial/sftime.

\bibitem[{Wang \emph{et~al.}(2020)Wang, Cook, and Hyndman}]{tsibble}
Wang E, Cook D, Hyndman RJ (2020).
\newblock \enquote{A New Tidy Data Structure to Support Exploration and Modeling of Temporal Data.}
\newblock \emph{Journal of Computational and Graphical Statistics}, \textbf{29}(3), 466--478.
\newblock \urlprefix\url{https://doi.org/10.1080/10618600.2019.1695624}.

\bibitem[{Wickham(2014)}]{tidydata}
Wickham H (2014).
\newblock \enquote{Tidy Data.}
\newblock \emph{Journal of Statistical Software}, \textbf{59}(10), 1–23.
\newblock \urlprefix\url{https://doi.org/10.18637/jss.v059.i10}.

\bibitem[{Wickham \emph{et~al.}(2012)Wickham, Hofmann, Wickham, and Cook}]{Wickham2012-yr}
Wickham H, Hofmann H, Wickham C, Cook D (2012).
\newblock \enquote{Glyph-Maps for Visually Exploring Temporal Patterns in Climate Data and Models.}
\newblock \emph{Environmetrics}, \textbf{23}(5), 382--393.
\newblock \doi{10.1002/env.2152}.

\bibitem[{Xie(2015)}]{knitr}
Xie Y (2015).
\newblock \emph{Dynamic Documents with \proglang{R} and \pkg{knitr}}.
\newblock 2nd edition. Chapman and Hall/CRC, Boca Raton, Florida.
\newblock ISBN 978-1498716963, \urlprefix\url{https://yihui.name/knitr/}.

\bibitem[{Xie \emph{et~al.}(2018)Xie, Allaire, and Grolemund}]{rmarkdown}
Xie Y, Allaire J, Grolemund G (2018).
\newblock \emph{\proglang{R} Markdown: The Definitive Guide}.
\newblock Chapman and Hall/CRC, Boca Raton, Florida.
\newblock ISBN 978-1138359338, \urlprefix\url{https://bookdown.org/yihui/rmarkdown}.

\bibitem[{Xie \emph{et~al.}(2014)Xie, Hofmann, and Cheng}]{xie2014reactive}
Xie Y, Hofmann H, Cheng X (2014).
\newblock \enquote{{Reactive Programming for Interactive Graphics}.}
\newblock \emph{Statistical Science}, \textbf{29}(2), 201 -- 213.
\newblock \urlprefix\url{https://doi.org/10.1214/14-STS477}.

\end{thebibliography}

\end{document}